# Reply to Comments of Steuernagel on the Afshar's Experiment

Eduardo V. Flores

Dept. of Physics and Astronomy, Rowan University, Glassboro, NJ

**Abstract**

We respond to criticism of our paper "Paradox in Wave-Particle Duality for Non-Perturbative Measurements". We disagree with Steuernagel's derivation of the visibility of the Afshar experiment. To calculate the fringe visibility, Steuernagel utilizes two different experimental situations, i.e. the wire grid in the pattern minima and in the pattern maxima. In our assessment, this procedure cannot lead to the correct result for the complementarity properties of a wave-particle in one particular experimental set-up.

## 1 Introduction

We reply to the recent comments on our paper [1] by Steuernagel [2] and thank them for helping us explain our experiment more clearly. We briefly summarize what we consider their most relevant criticism and our response to it.

Steuernagel questions our analysis of the experimental results of the Afshar experiment and states that their correct analysis does not pose a paradox to Bohr's Principle of Complementarity. We want to show errors in their analysis and point out that the paradox we propose [1] remains unsolved.

In summary, Steuernagel separates photons into two mutually exclusive subensembles to which the complementarity principle must therefore be applied separately. One ensemble is made of those photons that cross a wire grid and the other one is made of those that do not cross the wire grid. We argue that their claim that the two ensembles are mutually exclusive for the purpose of determination of the complementarity principle is unsupported. Steuernagel assumes the presence of an ideal interference pattern at the grid with visibility 1 and then shows that the measured visibility of this pattern is much less than 1. We argue that it is better not to assume a particular interference pattern and let the experimental results determine the possible pattern and its visibility. Finally, we point out that their use of two different experimental set-ups to obtain the visibility constitutes a major error in their work.

## 2 A Quick Introduction to the Afshar Experiment

The Afshar experiment consists of coherent light incident onto a pair of pinholes [1]. The two emerging beams from the pinholes spatially overlap in the far-field and interfere to produce a pattern of alternating light and dark fringes. At an appropriate distance from the pinholes thin wires are placed at the minima of the interference pattern. Beyond the wires there is a lens that forms the image of the pinholes onto two photon detectors located at the image of each pinhole. When an interference pattern is not present, as in the case when only one pinhole is open, the wire grid obstructs the beam and produces scattering, thus reducing the total flux at the corresponding detector by about 14%. However, when the interference pattern is present the disturbance to the incoming beams due to the wires is minimal, about 1%. From comparative measurements of the total flux with and without the wire grid, the presence of an interference pattern is inferred in a non-perturbative manner. Thus, the parameter V that measures the visibility of the interference pattern is near its maximum value of 1.

When the wire grid is not present quantum optics predicts that a photon that hits a given detector originates from the corresponding pinhole with a very high probability. The parameter K that measures the "which-way" information is 1 in this case. When a wire grid is placed at the dark fringes, where the wave-function is zero, the photon flux at the detectors hardly changes. We argue [1,3,4] that this is an indication that the wires have barely altered the "which-way" information, thus, K is also nearly 1 in violation of the Greenberger-YaSin inequality $V^2 + K^2 \leq 1$, a modern version of Bohr's principle of complementarity [5].

## 3 Our Calculation of the Visibility

A very small decrease in the photon count (of the order of 1%), when the wire grid is at a presumed region of destructive interference, is compatible with the presence of perfect two-pinhole interference pattern with visibility of 1. However, due to Heisenberg's uncertainty principle, we cannot measure its visibility directly without compromising the which-way information [6]. Fortunately, we can provide a lowest limit for the visibility compatible with our data.

When both pinholes are open the small decrease in the photon count (of the order of 1%) due to the presence of the wire grid can only be explained by destructive interference at the location of the wires. Thus, we have evidence for an interference pattern at the wire grid. We start by assuming ignorance about its shape. We seek for an interference pattern compatible with the data and with the lowest possible visibility to place a lowest limit.

We consider the standard formula for the visibility $V = \frac{I_{max} - I_{min}}{I_{max} + I_{min}}$, where $I_{max}$ and $I_{min}$ are the maximum and minimum intensities of the interference pattern. To minimize the visibility $I_{max}$ needs to be as small as possible and $I_{min}$ as large as possible.

We notice that $I_{min}$ must be found at the location of the wires to explain the small decrease in photon count. Not all the losses in photon count happen at the wires as there are photons that are diffracted to higher orders and do not reach detectors. Yet, to maximize $I_{min}$ we put as many photons as possible in the very small area covered by the wires. Thus, we place all the losses in photon count (1%) in $I_{min}$. Therefore, the regions with minimum intensity have the geometrical shape of a thin rectangular box of base equal to the thickness times the length of a wire. The height of the boxes is proportional to 1% divided by the area of the base.

The maximum intensity regions are minimized by distributing the photons that miss the wires in large boxes of base equal to the regions not covered by the wires and of height proportional to 99% over the area of the large base. Thus, the interference pattern with the lowest visibility is a type of periodic square function. A simple calculation using the data of the Afshar experiment gives the lowest limit for the visibility, $V \geq 0.64$ [1].

Steuernagel criticize our use of the 1% decrease in photon count at the detectors in the calculation of the visibility. They argue [2] that

> "transmitted and back-reflected (or absorbed) photons form two mutually exclusive subensembles to which the complementarity principle must therefore be

> applied separately—precisely because they are subjected to different simultaneous path and wave measurements."

However, in our experiment, all the photons are subject to the identical path and wave measurements. The fact that some end up at the detectors, some at the wires and some get diffracted to higher orders is not under our control. What is under our control is providing an identical environment for the photons before measurement.

We use a low photon rate of $3 \times 10^4$ photons/sec [1] so that the average separation between successive photons is about 10 km. We claim that there is a single particle (either a single photon or a single bunch of photons) in the set-up at any given time. Each particle that goes through the apparatus encounters an identical set-up, thus, each particle has exactly the same amplitude for a particular interaction with the set-up. The amplitude to go through the wire grid is the same for all particles. The amplitude of being absorbed or back-reflected by the wire grid is also the same for all particles. These amplitudes apply individually to each particle and together contribute to a calculation of its properties: which-way information, visibility, etc. We do not see a mathematical or physical reason to form two mutually exclusive subensembles of photons just because some particles end up at one place or another under identical conditions.

## 4 Steuernagel Calculation of Visibility

Steuernagel calculation [2] of the visibility starts by considering an ideal interference pattern at their grating (wire grid equivalent):

> "The light in the grating plane forms a sinusoidal field distribution, with the intensity distribution $I(x) = Cos(\pi / \Lambda x + \phi)^2$, where $\phi$ is the relative phase between the two slits $S_1$ and $S_2$. To find out how much light gets transmitted we have to integrate over the grating's slit opening(s). We find that the transmitted intensity is given by $I_{t,\max} = \int_{-\Lambda/2(1-a)}^{\Lambda/2(1-a)} dx \cos(\pi/\Lambda x)^2 = (\pi - a\pi + \sin(a\pi))/(2\pi)$ in the maximum case (grating positioned at interference pattern minima, $\phi = 0$) and

$I_{t,\min} = \int_{-\Lambda/2(1-a)}^{\Lambda/2(1-a)} dx \sin(\pi/\Lambda x)^2 = (\pi - a\pi - \sin(a\pi))/(2\pi)$ in the minimum case (grating positioned at interference pattern maxima, $\phi = \pm\pi$). The ensuing measurable visibility of transmitted light $V_t$ thus is $V_t = \dfrac{I_{t,\max} - I_{t,\min}}{I_{t,\max} + I_{t,\min}} = \dfrac{\sin(a\pi)}{\pi(1-a)}$."

For a numerical example of their visibility formula they use a grid with covering ratio $a = 0.05$ and get $V_t = 0.0524$. We notice that it is Steuernagel's calculation that requires two truly independent ensembles of photons. They are independent because they are obtained by different experimental set-ups. These are precisely the types of ensembles which should not be mixed for a calculation of complementarity properties. The first ensemble comes from a set-up with the wire grid at the minima of the interference pattern, $\phi = 0$. For the second ensemble the experimental set-up changes, the wire grid is repositioned at the maxima of the interference pattern, $\phi = \pm\pi$. We notice that following Steuernagel's approach we could as well in one experimental set-up place an opaque screen in front of the interference pattern and measure its visibility and in the second experimental set-up remove the screen and the wire grid, measure the which-way information and use these two independent values to calculate complementarity properties.

## 5 Conclusions

We find no support for Steuernagel's claim that our single experimental set-up produces two mutually exclusive photon ensembles for the purpose of calculating complementarity properties. We actually find that their criticism applies to their own approach: Steuernagel uses two different experimental set-ups to obtain the visibility. Even if they only intended to measure the visibility and not to take into account the simultaneous which-way information, their approach still seems ineffective. Notice that they started with an ideal pattern with theoretical visibility of 1 and ended up with a visibility value of 0.0524. Our calculation of the visibility does not assume a particular interference pattern but sets an experimental lowest limit for the visibility of any pattern actually present at the wire grid. Other criticisms are not addressed in this reply as they are minor compared with the calculation of the visibility.